\documentstyle[12pt,aaspp4]{article}

\slugcomment{Submitted to ApJ Letters/Revised}

\lefthead{Ohl et al.}
\righthead{UV Color Gradients}

\begin{document}

\title{Far-Ultraviolet Color Gradients in Early-Type Galaxies}

\author{R. G. Ohl\altaffilmark{1} and R. W. O'Connell}
\affil{Astronomy Department, University of Virginia, P.O. Box 3818,
    Charlottesville, VA 22903}

\author{R. C. Bohlin}
\affil{Space Telescope Science Institute,  
3700 San Martin Drive, Baltimore, MD 21218}

\author{N. R. Collins\altaffilmark{2}, B. Dorman\altaffilmark{3}, 
M. N. Fanelli\altaffilmark{2} and S. G. Neff}
\affil{Laboratory for Astronomy and Solar Physics, Code 681, 
NASA-Goddard Space Flight Center, Greenbelt, MD 20771}

\author{M. S. Roberts}
\affil{National Radio Astronomy Observatory,  
Edgemont Road, Charlottesville, VA 22903}

\and

\author{A. M. Smith and T. P. Stecher}
\affil{Laboratory for Astronomy and Solar Physics, Code 681, 
NASA-Goddard Space Flight Center, Greenbelt, MD 20771}

\altaffiltext{1}{Also Department of Physics and Astronomy, 
Johns Hopkins University, Charles and 34th Streets, Baltimore, MD 21218}

\altaffiltext{2}{Raytheon STX Corp., Lanham, MD 20706}

\altaffiltext{3}{Also Astronomy Department, University of Virginia, 
P.O. Box 3818, Charlottesville, VA 22903}

\begin{abstract}

We discuss far-UV (1500 \AA) surface photometry and FUV--B color
profiles for 8 E/S0 galaxies from images taken with the Ultraviolet
Imaging Telescope, primarily during the {\it Astro-2} mission.  In
three cases, the FUV radial profiles are more consistent with an
exponential than a de~Vaucouleurs function, but there is no other
evidence for the presence of a disk or of young, massive stars.  In
all cases except M32 the FUV--B color becomes redder at larger radii.
There is a wide range of internal radial FUV--B color gradients.
However, we find no correlation between the FUV--B color gradients and
internal metallicity gradients based on Mg absorption features.  We
conclude that metallicity is not the sole parameter controlling the
``UV upturn component'' in old populations.

\end{abstract}


\section{Introduction}

Radial photometric gradients in the old populations of elliptical
galaxies have been recognized for almost 40 years as fundamental clues
to galaxy evolution.  At optical wavelengths the amplitudes of
gradients are modest.  They are challenging to measure and are known
to suffer from a serious age-abundance degeneracy (e.g.\ O'Connell
1986; Worthey 1994), wherein it is difficult to distinguish small
changes in age from small changes in metallicity.  However,
ellipticals also contain a stellar component with an effective
temperature higher than 20000 K which appears to be unusually
sensitive to the age and chemical abundances of their old
populations.  The hot component is well isolated in the
far-ultraviolet (FUV, $\lambda < 2000$ \AA) from the cooler main
sequence and red giant branch components which dominate galaxy light
at longer wavelengths.  FUV observations therefore offer an important
means of assessing population characteristics independent of those
widely used to date, once the physics of the hot component is well
understood.  In this paper we discuss spatial gradients in the hot
component and their correlation with gradients in the optical-band Mg$_2$
metal line index.

The hot ``UV upturn'' or ``UVX'' component was first detected in 
FUV photometry of nearby spiral bulges and E/S0 galaxies by the OAO and
ANS experiments (Code \& Welch 1979, de
Boer 1982).  It was subsequently identified with the old populations
of these systems after alternative candidates, most prominently young
massive main sequence stars, were excluded through UV spectroscopy
(e.g.\ Oke, Bertola, \& Capaccioli 1981; Welch 1982; O'Connell, Thuan
\& Puschell 1986; Burstein et al.\ 1988; 
Ferguson \& Davidsen 1993; Brown et al.\ 1997) and UV imaging (Bohlin
et al.\ 1985; O'Connell et al.\ 1992; King et al.\ 1992; Bertola et
al.\ 1995; Cole et al.\ 1998).

The UVX varies considerably among and within galaxies.  FUV--V colors
exhibit a much larger range (2-3 mags) between galaxies than do the
more familiar optical-IR colors for old populations.  There is
superficially a good correlation with metal abundance:  the nuclear UV
colors of E galaxies, measured with the $10\arcsec \times 20\arcsec$
aperture of IUE, correlate well with the index Mg$_2$, which measures
the Mg~I $+$ MgH absorption features near 5170 \AA\ (Burstein et al.\
1988).  The FUV--V colors become bluer (i.e.\ the UVX becomes
stronger) as Mg$_2$ increases.  However, the available sample is not
large enough to distinguish a smooth relation between the UVX and Mg$_2$
from the effects of several distinct groups of systems (Dorman,
O'Connell, \& Rood 1995).  

We found strong radial gradients in UV colors within galaxies from
photometry with the Ultraviolet Imaging Telescope (UIT) obtained
during the 1990 {\it Astro-1} mission (O'Connell et al.\ 1992).  The
measurements, at 1500\AA\ and 2500\AA, covered the inner
1\arcmin\--2\arcmin\ of M31 (Sb), M81 (Sb), NGC$\,$1399 (gE), and M32
(dE).  In the first three cases, the UVX became stronger at smaller
radii as line strengths increased.  M32 exhibited a large reversed
gradient, with the UVX becoming stronger at larger radii (up to $r \sim
30\arcsec$), even though its optical color gradients are
small (Davidge 1991, Peletier 1993, Hardy et al.\ 1994).  

The existence of large and easily measured UV color gradients
encouraged us to observe a larger sample of E/S0 galaxies during the
1995 {\it Astro-2} mission.  Here we report on the brighter objects in
the {\it Astro-2} sample and examine correlations between the radial
gradients in the UVX and Mg line strengths.

\section{Observations and Results}

The galaxy sample included here spans a large range of absolute
magnitude and metallic line strengths.  In order of decreasing central
UVX strength (\cite{burstein88}), the galaxies with their Revised
Shapley Ames Catalog morphological classifications (Sandage \& Tammann
1987) are M87 [E0], NGC$\,$1399 [E1], M60 [S01(2)], M89 [S01(0)], M49
[E1/S01(1)], NGC$\,$3115 [S01(7)], NGC$\,$3379 [E0] and M32 [E2].  All
UV data are from the March 1995 {\it Astro-2} mission with the exception of
NGC$\,$1399, for which we have re-reduced the {\it Astro-1} data using
the final flux calibration for that mission.  

The UIT is a $f/9$ Ritchey-Chr\'{e}tien telescope with a 38 cm diameter
clear aperture.  It offers a large field of view ($40'$ diameter) and
good spatial resolution ($\rm{FWHM} \sim 3''$), allowing detailed
two-dimensional UV photometry of galaxies.  Only images with the FUV
filters B1 and B5 and the longest exposure times (550--1560~s) are
analyzed here.  These have effective wavelengths of 1520 and 1615
\AA\ and full widths at half maximum of 350 and 225 \AA, respectively.
The detector is a two-stage image intensifier coupled with fiber
optics to Kodak~IIa-O film.  The CsI photocathode provides excellent
rejection of longer-wavelength photons, and no ``red leaks'' are
detectable in observations of cool systems like E/S0 galaxies.
Details of the basic data reduction are given in Stecher et al.\ (1997).
We employ the final flux calibrations from {\it Astro-1}
and {\it Astro-2} here, denoted ``FLIGHT15'' and ``FLIGHT22,''
respectively, the latter based on photometry of 49 stars observed in
common with IUE.  We confirmed that the stellar calibration is valid
for extended sources by comparing {\it Astro-2} data of the
bright reflection nebula NGC$\,$7023 (Witt et al.\ 1992) with NEWSIPS
(Nichols \& Linsky 1996) IUE spectra of this object.  The overall flux
zero-point of UIT photometry is accurate to about 10\%.  In this paper
we use the monochromatic magnitude system for the FUV, where $
m_{\lambda} = -2.5 \log
\rm{F}_{\lambda} - 21.1 $ and $
\rm{F}_{\lambda} $ is the incident flux in units of $
\rm{ergs\, s^{-1}\, cm^{-2}\, \AA^{-1}} $.

Galaxy FUV photometry was performed as in O'Connell et al.\ (1992),
using circular annuli.  Sky background levels were generally
determined from photometry in large annuli centered on each galaxy.
Typical sky brightnesses are $\mathrm{ \mu (1500 \AA)} \ga 23.1$ mag
$\mathrm{arcsec^{-2}}$.  Because the UIT mid-UV camera failed early in
the {\it Astro-2} mission, we cannot measure the (1500--2500 \AA)
colors we used in O'Connell et al.\ (1992).  Instead, we have compared
our FUV surface photometry to  B-band photometry from 
\cite{michard85} for M89; \cite{peletier90} for M49, M60, M87, and
NGC$\,$3379; \cite{peletier93} for M32; Cheng et al.\ (1996) for
NGC$\,$1399; and our own CCD observations for NGC$\,$3115.  Optical
surface brightness measurements made along elliptical isophotes
with semi-major axis $a$ and semi-minor axis $b$ were projected to the
intermediate radius $\mathrm{\sqrt{a \times b}}$ for comparison with
FUV measurements made with circular annuli.  Most galaxies in our
sample have axial ratios near 1.0, and numerical experiments showed
that the FUV--B color profiles are not biased by this process.  Since we
are interested only in the shapes of the profiles, the data in the
figures have not been corrected for foreground Galactic extinction.  

In agreement with earlier studies (Bohlin et al.\ 1985, O'Connell et
al.\ 1992, Bertola et al.\ 1995, Cole et al. 1998), we find the FUV
emission from most of these objects to be distributed smoothly on the
sky, not in clumps or knots, as would be expected if the UV light was
produced by young, massive stars (the FUV images will appear in Marcum
et al., in preparation).  In general, the FUV contours of our sample
are consistent in shape and orientation with B-band isophotes.  There
is no UV morphological evidence for significant massive star formation
in these galaxies. 

M87 and NGC$\,$3379 are the only objects showing FUV emission that is
not smooth on the sky.  In the case of M87, we see the UV counterparts
of the bright active nucleus and non-thermal jet.  The FUV structure
in the jet is consistent with that observed at other wavelengths (e.g.
\cite{sparks96}).  The nucleus and jet were excluded from the analysis
presented in Figures~1-3, even though they contribute only about 17\%
of the FUV light within 20\arcsec.  There is also some
optical-band evidence that M87 contains a young-star ``accretion
population'' associated with its X-ray cooling flow (McNamara \&
O'Connell 1989), which may produce its unusually flat 2000-2800 \AA\
spectrum (Burstein et al.\ 1988).  This could account for the modest
differences in its FUV profile compared to the rest of our sample (see
below); however, its gross FUV properties are similar to the other
objects.  

NGC$\,$3379 shows FUV evidence of a dust lane or other dark linear
structure $\sim 10\arcsec$ in length to the SE of the nucleus at
$\mathrm{r \sim 11\arcsec}$.  Its effects are visible in the profiles plotted
in Figures~1 and 2.  This feature is distinct from the small ($r \sim \,
$1.4\arcsec) ring discussed by van Dokkum \& Franx (1995) and has not
been reported at other wavelengths.  

The B-band profiles of all eight objects are well fitted by $r^{1/4}$
de~Vaucouleurs functions, which are characteristic of spheroids.  In
the FUV, M89, NGC$\,$3115, and M32 also display roughly de~Vaucouleurs
profiles (see Figure~1), as does NGC$\,$1399 in both the FUV and mid-UV
(\cite{oconnell92}).  By contrast, the inner FUV profiles
for M60, M49, and NGC$\,$3379 are more consistent with an exponential
function, though the profiles tend to flatten at large radii.  M87's
FUV profile is consistent with neither function.  Because of the
significant FUV--B color gradients, it is not necessarily expected
that the UV profiles of objects which are true spheroids at optical
wavelengths would be strictly de~Vaucouleurs in shape.  What is
perhaps more surprising is that some appear to be exponential.  Although
exponentials are normally associated with disks, the FUV isophotal
contours are consistent with the B-band contours in shape and
orientation, and the 3-dimensional FUV light distributions are
therefore unlikely to be genuinely disklike.   

The FUV--B color profiles are plotted in Figure~2.  The objects
display a large (3 mag) range in central colors.  This is very
different from the behavior of early-type galaxies at optical
wavelengths (e.g.\ Peletier et al.\ 1990).  The color ranking of the
objects is consistent with that based on IUE spectroscopy
(\cite{burstein88}).  Five of the galaxies (M87, NGC$\,$1399, M60,
M49, and M32) display large internal FUV--B color gradients, with net
changes larger than 1.0 mag over the region photometered.  Again, this
is unlike the very mild color gradients encountered in the optical and
IR.  M89, NGC$\,$3115, and NGC$\,$3379 have smaller overall color
changes.  M32 is the only object which becomes bluer in FUV--B at
larger radii, confirming our {\it Astro-1} photometry (O'Connell et
al.\ 1992).  The sense of the FUV color gradients inferred by Brown et
al.\ (1997) within six of these objects by comparison between IUE
spectra ($10'' \times 20''$ aperture) and {\it Astro-2}/HUT spectra
($10'' \times 56''$) is also consistent with our higher resolution
photometry.  Internal extinction by dust cannot be responsible for
these gradients.  They are so large that significant optical-band
effects would be expected, since $A(4400{\rm \AA}) \sim 0.5
A(1500$\AA), where $A$ is the total extinction in magnitudes.  {\it
Astro}/HUT spectroscopy also places strict, and low, limits on the
amount of internal extinction (Ferguson \& Davidsen 1993, Brown et
al.\ 1997). 

Interestingly, all the objects except M32 exhibit an inner ``plateau''
in color before a steeper reddening sets in, usually at radius $r
\simeq 20''$.  There is almost a 2-component structure in some cases.
The outermost measures have lower precision, but the breaks in the mean
gradients are well determined.  Such structure is not common in
optical data, though overall amplitudes are lower there, which could
prevent its detection in some cases.  

The mean logarithmic gradient ($\Delta\, {\rm Color}/ \Delta\, \log r$) in
FUV--B for each object was determined by a  fit to the FUV--B color profile
for the region interior to the optical half-light radius (de
Vaucouleurs et al.\ 1976).  Data points were weighted by the
inverse of their estimated photometric uncertainties, 
so the gradients are representative of the inner parts of the
profiles.  This method is consistent with that used by Gonz\'{a}lez
\& Gorgas (1996) for the mean $\mathrm{Mg_{2}}$ gradient.

In Figure~3 we plot the internal FUV--B color gradient against the
internal Mg$_2$ gradient.  Mg$_2$ gradients for M32, M49, M60, M89,
NGC$\,$3115, and NGC$\,$3379 were kindly provided by J.J. Gonz\'{a}lez
(\cite{gonzalez96}); H. Kuntschner kindly provided Mg$_2$ data for
NGC$\,$1399; the gradient for M87 is from Davies, Sadler \& Peletier
(1993).  There is no correlation evident between the two gradients.
Although the sample is small, it spans the range of observed internal
gradients in both UV and metallicity properties.

\section{Discussion}

The best evidence now is that the UVX originates predominantly from
He-burning, extreme horizontal branch (EHB) and post-HB evolutionary
phases.  These objects have very small envelope masses ($\la 0.04
M_{\odot}$) on the HB, and they are probably descendents of the
dominant metal-rich stellar population, rather than a metal-poor
minority component.  No more than $\sim 15$\% of the metal rich red
giants need pass through the small-envelope channel to produce the
strongest UV upturns.  Net UV output is governed mainly by mass loss
processes on the red giant branch and by the helium abundance.  Metal
abundance and age have a less direct effect, though these do combine
to influence mass loss.  This interpretation is based on theory and
IUE spectroscopy (e.g.\ Greggio \& Renzini 1990; Horch, Demarque, \&
Pinsonneault 1992; Bressan, Chiosi, \& Fagotto 1994; Dorman et al.\
1995; and Yi, Demarque,
\& Oemler 1997) and also on 912-1600 \AA\ spectra for E galaxies
obtained by the {\it Astro}/HUT experiment (Ferguson \& Davidsen 1993;
Brown et al.\ 1997).  Contributions by hot metal-poor stars, an
alternative urged most recently by Park \& Lee (1997), are small,
based on the shape of the spectra for 912--2500\AA\ (Dorman et al.\
1995; Brown et al.\ 1997; Yi, Demarque \& Oemler 1998).

The simplest interpretation of the UV color variations is then that
they are related to metallicity, probably in two ways.  First,
increased metallicity probably drives larger mass loss on the RGB,
producing HB objects with smaller envelopes and hence larger UV
output.  Second, it is possible that systems
with higher metallicities also have larger He abundances (with
$\Delta Y / \Delta Z \simeq 2$-3; e.g.\  Greggio \& Renzini 1990),
which yield larger UV outputs for a given envelope mass.  Recent
studies of Fe and Mg line indices (e.g.\ Worthey et al.~1992; Worthey
1998; Trager et al.\ 1998) demonstrate that although there are significant
variations in [Mg/Fe] among different elliptical galaxies, nonetheless
within a given galaxy [Mg/Fe] $\simeq$ {\it constant}.  Since the
internal Mg line strength gradients reflect metallicity gradients, we
then expect objects with larger Mg gradients to have larger UV color
gradients.  
 
Despite this expectation and the strong apparent correlation between
nuclear measures of the UVX and Mg (Burstein et al.\ 1988), the
direct comparison of FUV--B and Mg gradients in Figure~3 shows no
correlation.  The breaks in the color slopes shown in Figure~2, which
are not characteristic of optical gradients, also suggest a distinct
behavior for the UVX.  To study the latter further, we are
comparing local values of FUV--B and Mg$_2$ as a function of radius
(Ohl et al., in preparation).  

Our results then imply that the UVX may not be simply related to
overall metal abundances in galaxies and that other parameters are
involved.  Age is perhaps the leading candidate.  Other things
remaining constant, HB envelope mass should decrease, and the UVX
should increase, with increasing age.  M32, with a large and reversed
UVX gradient, is an important case.  It has optical gradients which
are smaller but in the same sense as the other objects.  It is notable
for the presence of an intermediate age ($\la 8$ Gyr) stellar
population and possibly an age gradient in which the central regions
are younger (O'Connell 1986, Freedman 1992, Rose 1994, Hardy et
al.~1994, Faber et al.\ 1995, Grillmair et al. 1996).  Its unusual UV
behavior supports the notion that age is important to the UVX.  

One alternative interpretation is that helium abundance is decoupled
from metallicity (i.e.\ $\Delta Y / \Delta Z \not=$ {\it constant}) in
elliptical galaxies.  There are few other observables in old
populations which are sensitive to He abundance, so the latter is an
intriguing possibility.

\acknowledgments

We are grateful to all the participants in the UIT and {\it Astro-2}
projects.  We thank J. J. Gonz\'{a}lez and H. Kuntschner for providing metallicity
gradient data in advance of publication and K. P. Cheng and P. M.
Marcum for providing ground-based optical images.  This work was supported
in part by NASA grants NAG5-700 and NAG5-6403 to the University of Virginia.

\clearpage

\figcaption[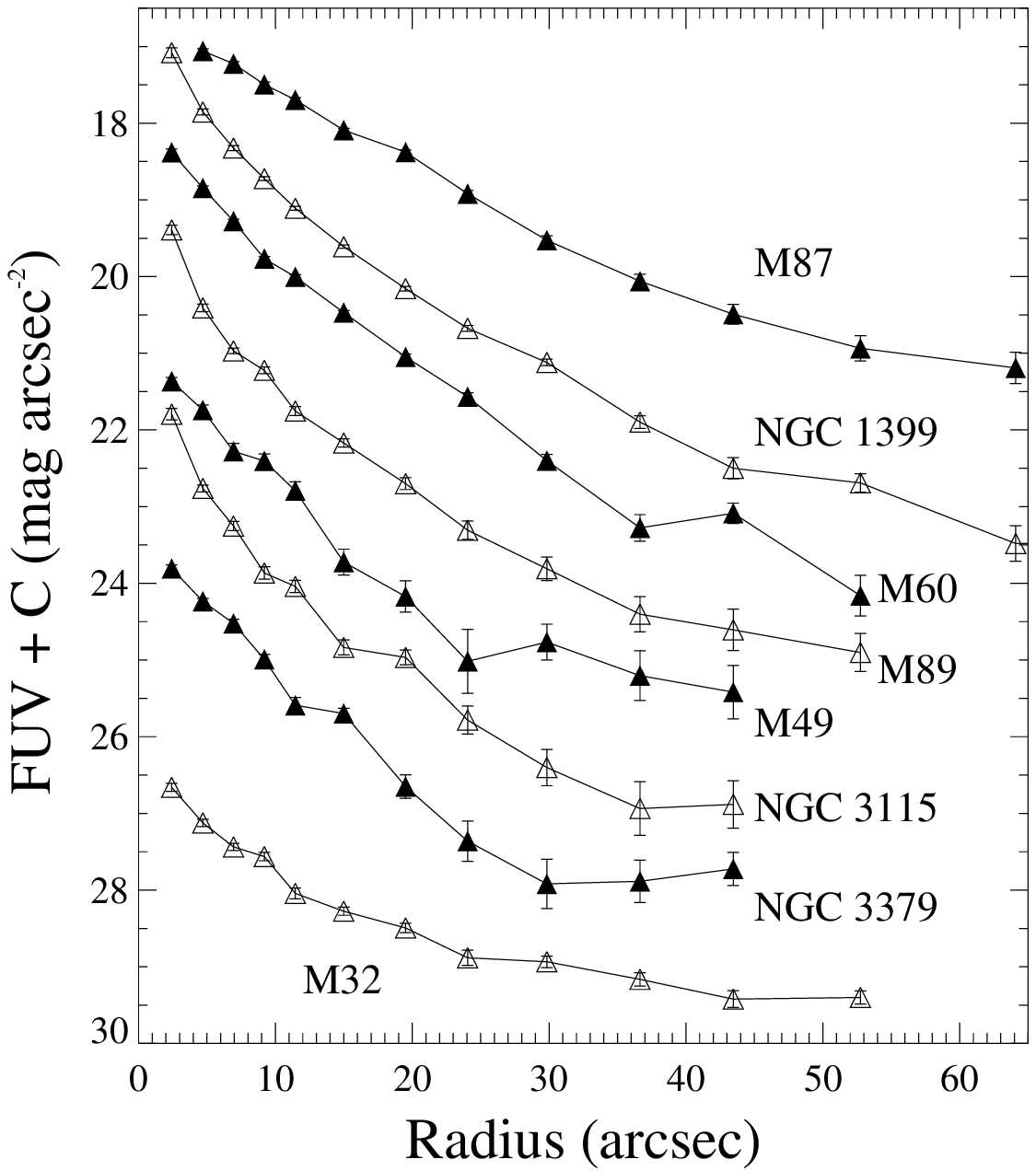]{FUV surface brightness profiles for eight
E/S0 galaxies.  One sigma error bars are shown.  In these units,
exponential profiles are straight lines.  For clarity, the profiles
are given zero-point offsets; from the top down these 
are $\mathrm{C=-3.0}$, $\mathrm{-2.0}$, $\mathrm{-1.0}$,
$\mathrm{0.0}$, $\mathrm{+1.0}$, $\mathrm{+2.0}$, $\mathrm{+3.0}$,
and $\mathrm{+6.5}$ mag $\mathrm{arcsec^{-2}}$.
\label{fig1}}

\figcaption[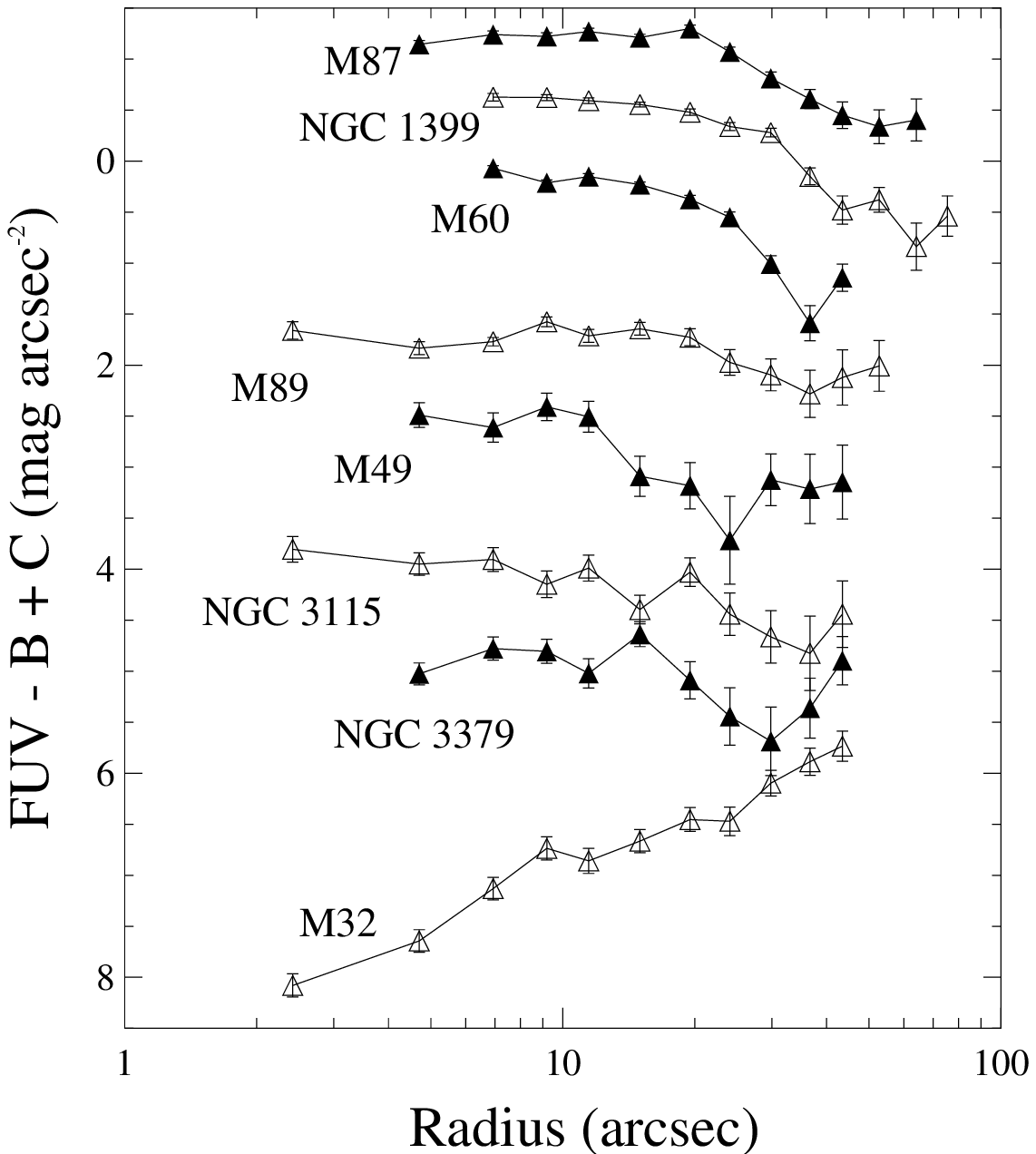]{Radial FUV--B color profiles for the
sample, offset for clarity and arranged in order of decreasing
central UVX.  One sigma error bars are shown. FUV--B
colors redden with increasing radius in all cases except M32.
Note the two-component structure in most of the profiles.   Offsets
in order from the top down are $\mathrm{C = -2.5}$, 
$\mathrm{-2.0}$, $\mathrm{-1.5}$, $\mathrm{0.0}$, $\mathrm{0.0}$, 
$\mathrm{+1.0}$, $\mathrm{+2.0}$, and $\mathrm{+3.5}$ mag 
$\mathrm{arcsec^{-2}}$. \label{fig2}}

\figcaption[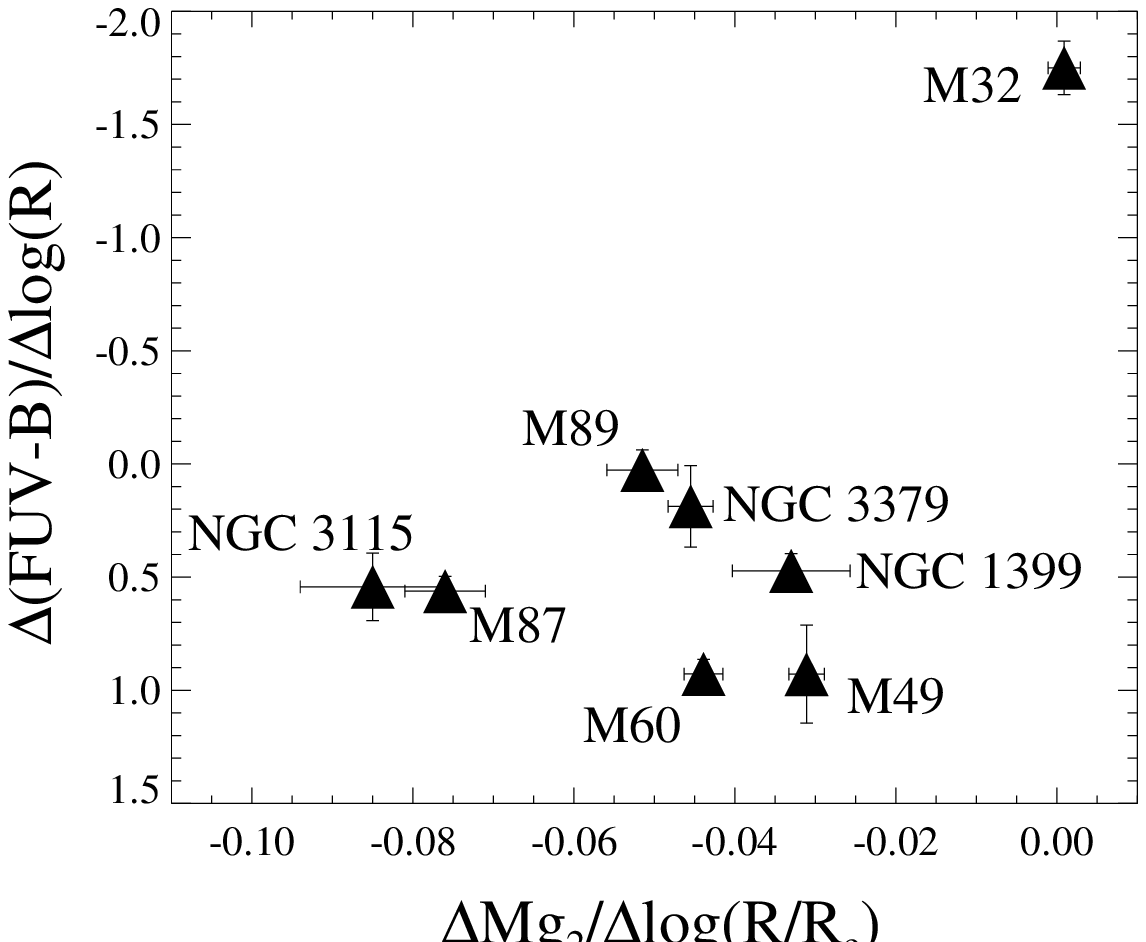]{The mean FUV--B internal color gradient versus
the mean internal Mg$_{2}$ line index gradient for the
sample.  One sigma error bars in the least-squares determination of
the FUV-B slope are shown. \label{fig3}}

\end{document}